\documentclass[showpacs,prb,aps,amsmath,amssymb,floatfix,twocolumn,floats]{revtex4}
\usepackage{color}
\usepackage{epsfig}
\usepackage{graphicx}
\begin{document}

\title{Spin dynamics of a tetrahedral cluster magnet}
\author{Wolfram Brenig}
\affiliation{Institut f\"ur Theoretische Physik, 
Technische Universit\"at Braunschweig, 38106 Braunschweig, Germany}

\date{\today}

\begin{abstract}
\begin{center}
\parbox{14cm}{
We study the magnetism of a lattice of coupled tetrahedral spin-1/2
clusters which might be of relevance to the tellurate compounds
Cu$_2$Te$_2$O$_5$X$_2$, with X=Cl, Br. 
Using the flow equation method we perform a series expansion in
terms of the inter-tetrahedral exchange couplings starting from the
quadrumer limit. Results will be given for the magnetic instabilities of
the quadrumer phase and the dispersion of elementary triplet excitations.
In limiting cases of our model of one- or two dimensional character we
show our results to be consistent with findings on previously investigated
decoupled tetrahedral chains and the Heisenberg model on the 1/5-depleted
square lattice.}
\end{center}
\end{abstract}


\maketitle

\section{Introduction}
\label{sec1}
Unconventional magnetism of frustrated spin systems has received
considerable interest recently.
Prominent examples are the one-dimensional (1D) frustrated spin-Peierls
compound CuGeO$_3$\cite{Hase93a} or the 2D orthogonal spin-dimer system
SrCu$_2$(BO$_3$)$_2$ with frustrating inter-dimer couplings\cite{Kageyama00a}.
Apart from dimer-based structures, frustrated spin systems involving triangular
or tetrahedral units, e.g. the kagom\'e-, checkerboard-, or the
pyrochlore-lattices are a focus of current research. In
the classical limit frustration leads to ground states with
macroscopic degeneracy in these
systems\cite{Ramirez96a,Ramirez00a}.
In the quantum case low-lying singlets seem to exist both,
on the kagom\'e and the checkerboard lattice with no long-range magnetic
order in the former\cite{Ramirez00a,Mila00a}, and a valence-bond-crystal
(VBC) ground state in the latter case\cite{Fouet01a,Brenig02a}. Analysis
of the 3D pyrochlore quantum-magnet remains an open issue\cite{Kawakami00a}.

\begin{figure}[bt]
\center{\psfig{file=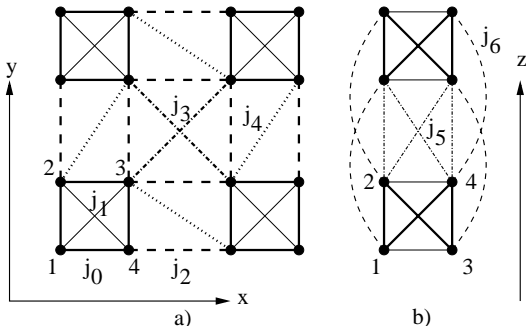,width=0.8\columnwidth}}
\caption{a) $xy$-plane and b) $z$-axis structure of the 3D
tetrahedral cluster-lattice. Spin-$1/2$ moments are
located on dotted vertices. $SU(2)$ type of exchange with
strength $j_{0,...,6}$ along the links.}
\label{fig2}
\end{figure}

Recently tellurate compounds, Cu$_2$Te$_2$O$_5$X$_2$, with X = Cl, Br
have been found to realize a new class of spin-1/2 systems where tetrahedra
of Cu$^{2+}$ align in tubes along the c(z)-direction and are separated
by lone-pair cations in the $ab(xy)$-plane \cite{Johnsson00a}.
Both, the effective dimensionality of this system as well
as the relevant magnetic interactions remain a puzzle.
Early analysis of thermodynamic data\cite{Johnsson00a,Lemmens01a}
was based on the 0D limit of isolated tetrahedral units,
i.e. on the exchange pattern of fig. \ref{fig2} with $j_{2,...6}=0$.
This resulted in $j_0=38.5(43)K$ and $j_1/j_0\sim 1$ for the
Chlorine(Bromide)
system which has been refined recently into $j_0\approx
47.66K$ and $j_1/j_0\approx 0.66$ for the Bromide system and 
$j_1/j_0<0.66$ for the Chlorine case\cite{Gros02a}. 
Raman spectroscopy\cite{Lemmens01a} however, indicates a
substantial inter-tetrahedral c-axis coupling.
This has prompted studies of 1D tet\-ra\-he\-dral
spin-chains\cite{Brenig01a,Totsuka02a} as in fig. \ref{fig2} with
$j_{2,3,4,6}=0$.
Yet, LDA calculations have given evidence of an additional z-axis
exchange path $j_6$ and transverse
inter-chain couplings as shown in \ref{fig2} a) of a magnitude which
can not be neglected\cite{Lemmens01a}.
In fact, specific heat data reveals a transition at $T_C=18.2(11.4)K$ in
the Chlorine(Bromide) system. In the Chlorine case the entropy change is
consistent with 3D antiferromagnetic (AFM) ordering. 

Combining fig. \ref{fig2} a) and b) a 3D cluster-spin model arises,
about which very little is known. We believe this to be an
interesting and highly frustrated magnetic system which
deserves to be investigated. Therefore,
the aim of this work is to shed light onto its excitations and possible
magnetic instabilities.  In addition, our analysis could be of relevance for the
Cu$_2$Te$_2$O$_5$X$_2$ system, in particular, if additional spectroscopic
data becomes available. In the remainder of this paper we first discuss our
method of calculation and then present results on the stability and the
triplet dynamics.

\section{Series Expansion}
\label{sec2}

The Hamiltonian, as read off from fig. \ref{fig2} can split into
a bare part $H_0$ and a perturbation $H_1$
\begin{eqnarray}\label{w1}
H_{\phantom 0} & = & H_{0}+H_{1} \nonumber \\
H_{0} & = & \sum _{{\bf l}}j_{0}({\bf S}_{1{\bf l}}+
{\bf S}_{3{\bf l}})({\bf S}_{2{\bf l}}+
{\bf S}_{4{\bf l}}) \nonumber \\
H_{1} & = & \sum _{{\bf l}}\, [\, j_{1}({\bf S}_{1{\bf l}}
{\bf S}_{3{\bf l}}+{\bf S}_{2{\bf l}}
{\bf S}_{4{\bf l}}) \nonumber \\
 &  & +j_{2}({\bf S}_{4{\bf l}}{\bf S}_{1{\bf l}+
 {\bf x}}+{\bf S}_{3{\bf l}}{\bf S}_{2{\bf l}+
 {\bf x}} \nonumber \\
 &  & \phantom {+j_{2}(}+{\bf S}_{2{\bf l}}{\bf S}_{1
 {\bf l}+{\bf y}}+{\bf S}_{3{\bf l}}{\bf S}_{4{\bf l}
 +{\bf y}}) + \nonumber
\end{eqnarray}

\begin{eqnarray}\label{w1a}
 &  & +j_{3}({\bf S}_{3{\bf l}}{\bf S}_{1{\bf l}+
 {\bf x}+{\bf y}}+{\bf S}_{2{\bf l}}
 {\bf S}_{4{\bf l}-{\bf x}+{\bf y}})\nonumber \\
 &  & +j_{4}({\bf S}_{3{\bf l}}{\bf S}_{1{\bf l}+
 {\bf x}}+{\bf S}_{2{\bf l}}{\bf S}_{4{\bf l}+
 {\bf y}})\nonumber \\
 &  & +j_{5}({\bf S}_{2{\bf l}}+{\bf S}_{4{\bf l}})
 ({\bf S}_{1{\bf l}+{\bf z}}+{\bf S}_{3{\bf l}+
 {\bf z}})\nonumber \\
 &  & +j_{6}\sum _{i}{\bf S}_{i{\bf l}}{\bf S}_{i{\bf l}
 +{\bf z}}\, ]\, \, \, ,
\end{eqnarray}

where the site of each tetrahedral unit is labeled by ${\bf l}$
with
${\bf S}_{i{\bf l}},\, i=1\ldots 4$ being the spin-$1/2$ operators
corresponding to each tetrahedron and \(
{\bf l}+{\bf x}({\bf y},{\bf z})$ refers to shifts of \(
{\bf l}$ by one unit cell along the $x,\, y$, or $z$-axis.

As has been pointed out previously \cite{Johnsson00a} the spectrum
of decoupled tetrahedra, i.e. for $j_{2,...6}=0$, is special in as
such that for $j_1<j_0$ each tetrahedral ground state is a singlet involving
all four spins, while at $j_1>j_0$ it is a product of two $S$=$0$-dimers
on each of the $j_1$-bonds. In turn, at $j_1=j_0$ the decoupled local ground
states are doubly degenerate singlets. This leads to quantum criticality
in the lattice case and bears the possibility of low-lying singlet
excitations\cite{Johnsson00a,Brenig01a,Totsuka02a}. This may be relevant
for the Bromide system but in the Chlorine compound $j_1$ seems
clearly less than $j_0$. Therefore in the remainder
of this paper we focus on the {\em quadrumer limit} of (\ref{w1}), defined
by setting $j_0\equiv 1$ and $j_{1,...6} << 1$.

The spectrum of each quadrumer consists of four {\em equidistant}
levels which can be labeled by spin $S$ and a number of local energy quanta
$q_{\bf l}$, c.f. table \ref{tab1}. The unperturbed Hamiltonian $H_0$ which
consists of the sum of quadrumers displays an equidistant ladder-spectrum
labeled by $Q=\sum_{{\bf l}} q_{\bf l}$. The $Q=0$ sector is the unperturbed
ground state $|0\rangle$ of $H_0$, which is a VBC
of quadrumer-singlets. The $Q=1$-sector contains local $S=1$
single-particle excitations of the VBC with
$q_{\bf l}=1$, where ${\bf l}$ runs over the lattice. At $Q=2$ the
spectrum of $H_0$ has total $S=0,1$, or $2$ and is of multi-particle nature.
For $S=0$ at $Q=2$ it comprises of one-particle singlets
with $q_{\bf l}=2$ and two-particle singlets constructed from triplets
with $q_{\bf l}=q_{{\bf m}}=1$ and ${\bf l}\neq {\bf m}$. The
perturbation $H_1$ in (\ref{w1}) can be written as a
sum of {\em two-site} operators $T_{n,k}$ which, for each coupling constant
$j_{k=1,...6}$ create(destroy) $n\geq 0$ ($n<0$) quanta within
the ladder spectrum of $H_0$.

\begin{eqnarray}
\label{w2}
H = H_0 + \sum^{N}_{n=-N}\sum^6_{k=1}j_k \, T_{n,k} 
\end{eqnarray}

It has been shown recently\cite{Stein97,Mielke98,Knetter00,Brenig02a}
that problems of type (\ref{w2}) allow for perturbative analysis
using a continuous unitary
transformation generated by the flow equation method of 
Wegner\cite{Wegner94}. The unitarily rotated effective
Hamiltonian $H_{\rm eff}$ reads\cite{Stein97,Knetter00}
\begin{eqnarray}
\label{w3}
H_{\rm eff}=H_0+\sum^\infty_{n=1}
\sum_{\stackrel{\mbox{\scriptsize $|{\bf m}|=n$}}{M({\bf m})=0}}
C({\bf m})\;W_{m_1}W_{m_2}\ldots W_{m_n}
\end{eqnarray}
where ${\bf m}=(m_1\ldots m_n)$ with $|{\bf m}|=n$ is an $n$-tuple of
integers, each in a range of $m_i\in\{0,\pm 1,\ldots,\pm N\}$ and
$W_{n}=\sum^6_{k=1}j_k T_{n,k}$. In contrast to $H$ of (\ref{w1}),
$H_{\rm eff}$ {\em conserves} the total number of quanta $Q$. This
is evident from the constraint $M({\bf m})=\sum^n_{i=1} m_i=0$.
The amplitudes $C({\bf m})$ are rational numbers computed from the
flow equation method \cite{Stein97,Knetter00}. Explicit
tabulation\cite{HttpRef} of the $T_{n,k}$ shows that for the Hamiltonian
in (\ref{w1}) $N=4$.

\begin{table}[t]
\begin{tabular}{|c|c|c|}
\hline
$E$ & $S$                 & $q_{\bf l}$ \\ \hline
1   & 2                   & 3           \\ \hline
0   & 0$\oplus$1$\oplus$1 & 2           \\ \hline
-1  & 1                   & 1           \\ \hline
-2  & 0                   & 0           \\ \hline
\end{tabular}
\caption{Energy ($E$, in units of $j_0$), spin ($S$), and 
quantum-number $q_{\bf l}$ of the quadrumer spectrum.}
\label{tab1} 
\end{table}

$Q$-conservation of $H_{\rm eff}$ leads to a ground state energy of
$E_g=\langle 0|H_{\rm eff}|0\rangle$. 
Evaluating this matrix element on clusters with periodic boundary conditions,
sufficiently large not to allow for wrap around at
graph-length $n$ one can obtain series expansions (SEs) for $E_g$ valid
to $O(n)$ in the thermodynamic limit, {\em i.e.} for systems of infinite size.
$Q$-conservation also guarantees the $Q=1$-triplets
remain genuine one-particle states. {\em A priori } single-particle 
states from sectors with $Q > 1$ will not only disperse via $H_{\rm eff}$, 
but can decay into multi-particle states. The dispersion of the
single-particle excitations is
\begin{eqnarray}
\label{eq5}
E_\mu({\bf k})=\sum_{lm} t_{\mu,lm} e^{i (k_x l+k_y m)}
\end{eqnarray}
where
$t_{\mu,lm}=\langle \mu,lm|H_{\rm eff}|\mu,00\rangle
- \delta_{lm,00}E^{obc}_g$
are hopping matrix elements from site $(0,0)$ to site
$(l,m)$ for a quadrumer excitation $\mu$
inserted into the unperturbed ground state.
For the thermodynamic limit $t_{\mu,lm}$ has to be evaluated on clusters with
open boundary conditions large enough to embed all linked
paths of length $n$ connecting sites $(0,0)$ to $(l,m)$ at $O(n)$ of
the perturbation. $E^{obc}_g=\langle 0|H_{\rm eff}|0\rangle$ on the
$t_{\mu,00}$-cluster. 

Previous applications of this method to spin systems were focused
on obtaining high-order SEs for one and two parameter
dimer \cite{Knetter00} and quadrumer \cite{Brenig02a} models in 1 or 2D. In
the present case, computational constraints related to the large number of
coupling constants and the 3D nature of the model confine the expansion to
4th-order. Moreover, explicit display of analytic expressions for
the elementary triplet dispersion has to be limited to 2nd order\cite{HttpRef}.

\begin{figure}[t]
\center{\psfig{file=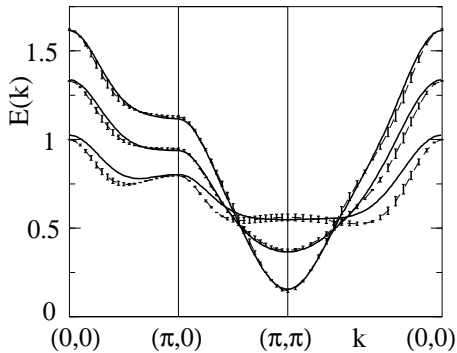,width=0.7\columnwidth}}
\caption{
Comparing the triplet dispersion along high-symmetry directions of a 2D
Brillouin zone as obtained from a 6th-order plaquette expansion
for the 1/5-depleted square lattice \protect{\cite{Gelfand96a}} (dashed with
error bars) with the 4th-order quadrumer expansion for the
tetrahedral spin-cluster model (solid) at $j_1=\gamma$,
$j_2=\lambda\gamma$, $j_4=\lambda$, and $j_{3,5,6}=0$, with $\lambda=1$ and
$\gamma=0.1$, $0.3$, and $0.5$ from top to bottom. (Upper and lower edges of error bars
refer to 4th- and 5th-order plaquette expansion.)}
\label{fig4}
\end{figure}

\section{Triplet excitations and magnetic instabilities}
\label{sec3}
In this section we analyze the triplet dispersion $E_{T}({\bf k})$ and the
stability of the quadrumer phase against magnetic ordering. We begin by
considering the result at 1st order in $j_{1,...6}$ for which we find
\begin{eqnarray}\label{eqn1}
E_{T}({\bf k}) & = & 1+\frac{1}{3}(j_{4}-2j_{2})(\cos (k_{x})+
\cos (k_{y}))\nonumber \\
 &  & +\frac{j_{3}}{3}(\cos (k_{x}+k_{y})+\cos (k_{x}-k_{y}))
 \nonumber \\
 &  & +\frac{4}{3}(j_{6}-j_{5})\cos (k_{z})
 \;\;.
\end{eqnarray}
Interestingly, this expression depends on three, effective
exchange coupling-constants only, i.e. $a=(j_{4}-2j_{2})/3$,
$b=j_{3}/3$, and $c=4(j_{6}-j_{5})/3$. Moreover $E_{{\bf k}}$ is
independent of $j_1$. In fact we find the dispersion to depend on
$j_1$ starting only at 3rd order. Due to the competition of the exchange
interactions in (\ref{w1}) the effective triplet hopping amplitudes $a$ and
$b$ in (\ref{eqn1}) can be of either sign, even for purely AFM $j_{2,4,5,6}$.

It is instructive to link (\ref{eqn1}) to other analytic results known from
related models. In particular, setting $j_1=j_2$ and $j_{3,5,6}=0$, the
tetrahedral cluster system of fig. \ref{fig2} is identical to a stack of
Heisenberg models on the 1/5-depleted square lattice\cite{Ueda96}.
Bond operator
theory (BOT) has been applied to this model yielding a triplet dispersion
of $E_{T}({\bf k}) = [1 + 2(j_4 - 2 j_2)/3 (\cos(k_x) + \cos(k_y))]^{1/2}$
in the quadrumer phase \cite{Starykh96a}. To 1st order this is obviously
identical to (\ref{eqn1}) with the same setting of parameters. Similarly,
for $j_{2,3,4,6}=0$ the quadrumer limit of fig. \ref{fig2} maps onto the
'dimerized spin-1 chain sector' of the tetrahedral-chain model  of
the tellurates studied in ref. \cite{Brenig01a,Totsuka02a}. BOT has been
applied also to that model, leading to
$E_{T}({\bf k}) = [1 - 8 j_5/3 \cos(k_z)]^{1/2}$.
Again, the latter is identical to 1st order with (\ref{eqn1}) with the same
choice of parameters. While this serves as a consistency check for
the series expansion we note that BOT, which is approximate only,
differs from the exact series already at 2nd order. Additional details on this
can be found in the appendix.

To test the quality of our perturbative expansion at 4th order we compare
to the plaquette series-expansion of Gelfand and collaborators
for the 1/5-depleted square lattice\cite{Gelfand96a}. This is achieved by
restricting the parameters in (\ref{w1}) to those of ref.\cite{Gelfand96a}, i.e.
$j_1=\gamma$, $j_2=\lambda\gamma$,
$j_4=\lambda$, and $j_{3,5,6}=0$. The plaquette series is
a {\em one} parameter expansion for a 2D model, which
allows for expansion of $E_{T}({\bf k})$ up to 6th order with
respect to $\lambda$, where the unperturbed Hamiltonian incorporates
$\gamma$ exactly. In fig. \ref{fig4} we contrast the 4th order results from
our {\em six} parameter expansion for the 3D tetrahedral spin system
with the plaquette expansion by considering the triplet dispersion. Despite
small deviations which set in upon increasing $\gamma$ the overall
agreement is satisfying. Since $\gamma$ is treated exactly within
the plaquette expansion, our 4th-order
quadrumer series does not coincide with one of the edges of the error bars
in fig. \ref{fig4} which refer to the 4th- and 5th-order plaquette series.  

\begin{figure}[b]
\center{\psfig{file=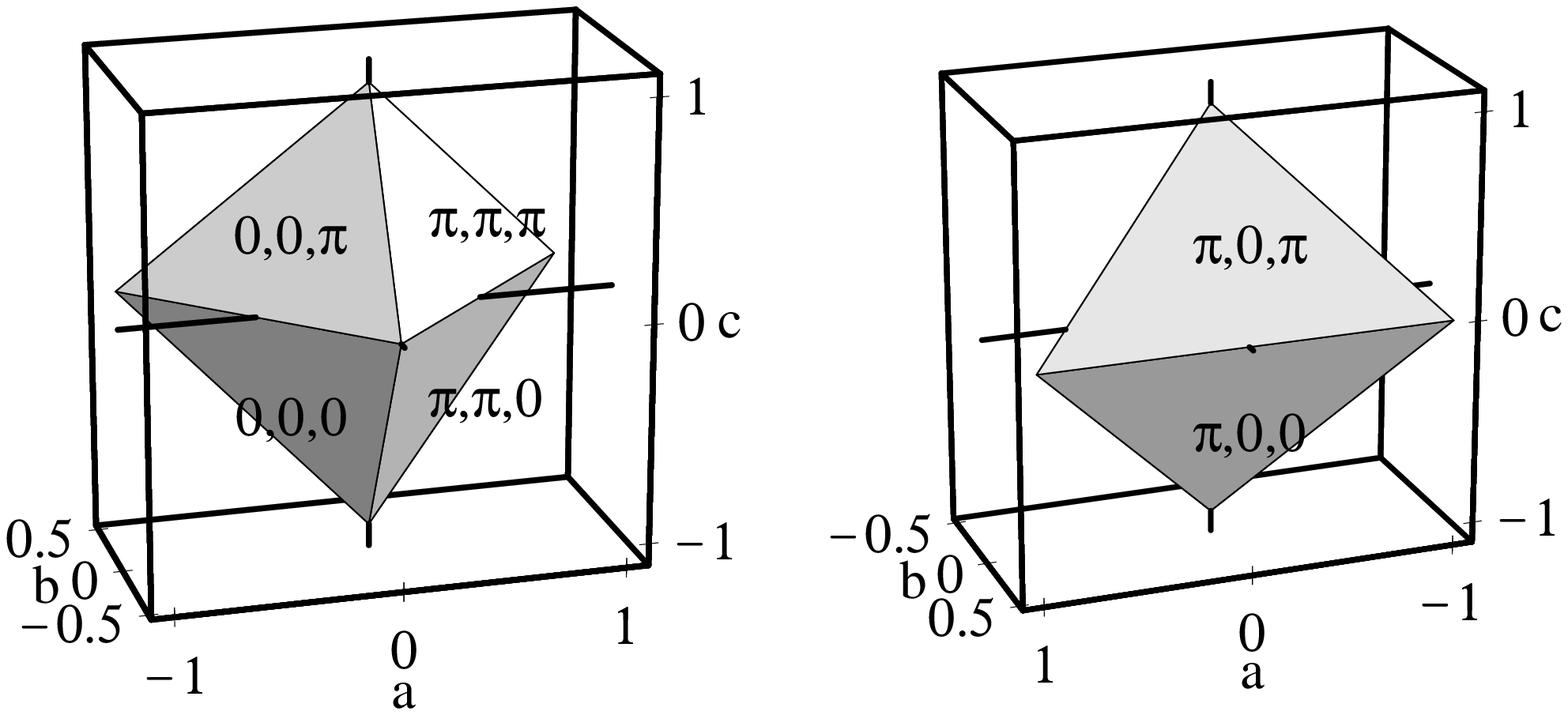,width=\columnwidth}}
\vspace{-.8cm}
\caption{Left(right) panel: rear(front) view of the stability surface
of the quadrumer phase at 1st order.
Faces are labeled by the wave vectors ${\bf k}_C$ of the instabilities
and $a=(j_{4}-2j_{2})/3$, $b=j_{3}/3$, and $c=4(j_{6}-j_{5})/3$.}
\label{fig5}
\end{figure}

Next we analyze the stability of the quadrumer phase against magnetic ordering
by identifying the surface in parameter space, closest to $j_{1,...6} = 0$
which  allows for triplet softening, i.e. the occurrence of a wave vector
${\bf k}_C$ with $E_{T}({\bf k}_C) = 0$. We emphasize that apart from such
instabilities, the tetrahedral spin system may exhibit others transitions, as
e.g. those related to the 1st-order, local quadrumer to dimer-product
transition on each of the tetrahedra. Here we focus on the triplet
softening only. To begin, in fig. \ref{fig5} we depict the instability
surface at 1st order as obtained from (\ref{eqn1}) along with the critical
wave vectors at which softening occurs. Due to the competing
interactions several ordering patterns are possible,
even for AFM couplings only.

Since the tetrahedral cluster-model contains six exchange coupling parameters
we will simplify the stability analysis at 4th order by selecting a subset of
them only. This selection is based on the effective exchange constants at 1st
order, i.e. we will focus on the stability as a function of $a$, $b$, and $c$
setting $j_{1,4,6}=0$. Figure \ref{fig6} a) shows the corresponding instability
surface. It has been obtained from a numerical search for zeros of the gap of
the 4th-order triplet-dispersion on a mesh of $21\times 41\times 2$ points
in the $abc$-space. For this purpose we
have used the bare series with no Pad\'e approximations applied. The surface is
not closed at its extremal extensions in the $ab$-plane, rather the stability
analysis has been confined to the range of parameters shown in this figure
in order to comply with the finite range of convergence of the perturbative
result. Only commensurate instability wave-vectors have been found within
the range of parameters investigated. The type of these wave vectors is
shown in \ref{fig6} b). While its shape is deformed with a reduced volume,
the main features of the 4th-order instability surface are still consistent
with those at 1st order. We find that the additional critical wave-vector types with appear along the
'edge-regions', i.e. at $m_{4,5}$ and $p_{4,5}$
occur within a parameter-range of poor convergence
of the perturbation theory. Therefore, these may be subject to
change at higher orders.

\setlength{\unitlength}{1pt}
\begin{figure}[t]
\begin{picture}(240,285)
\put(-12,11){\psfig{file=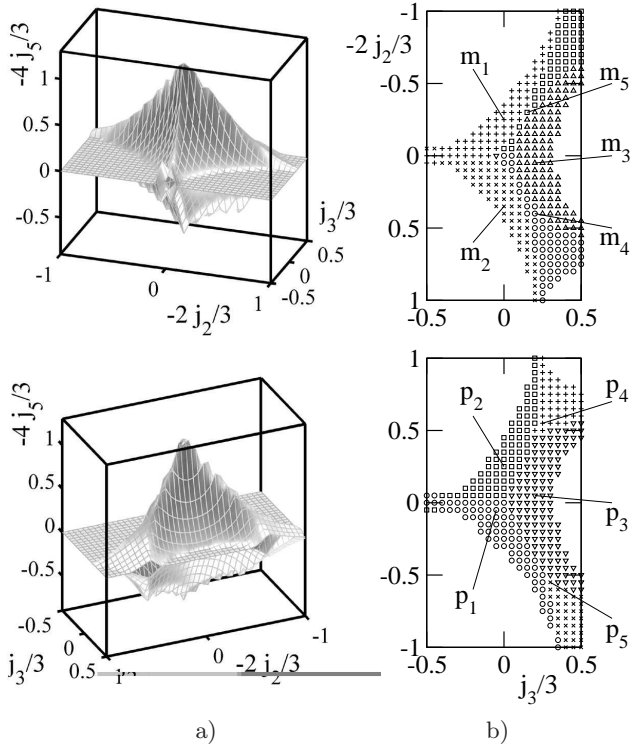,width=0.63\columnwidth}}
\put(130,9){\psfig{file=typsol.eps,width=0.44\columnwidth}}
\put(70,-3){ a)}
\put(180,-3){ b)}
\end{picture}
\caption{a) top(bottom) panel: rear(front) view of the stability surface
of the quadrumer phase at 4th order.
b) top(bottom) panel: wave vectors of instability for $j_5<(>)0$.
Labels refer to ${\bf k}_C$: 
$+$ =($0,0,\pi$), 
$\square$ =($\pi,\pi,0$), 
$\bigtriangleup$ =($\pi,0,\pi$), 
$\bigtriangledown$ =($\pi,0,0$), 
$\bigcirc$ =($0,0,0$),
and {\large $\times$} =($\pi,\pi,\pi$). Selected points $m(p)_{1,...5}$ refer
to fig. \protect{\ref{fig7}}}
\label{fig6}
\vskip-10pt
\end{figure}

\begin{figure}[t]
\center{
\psfig{file=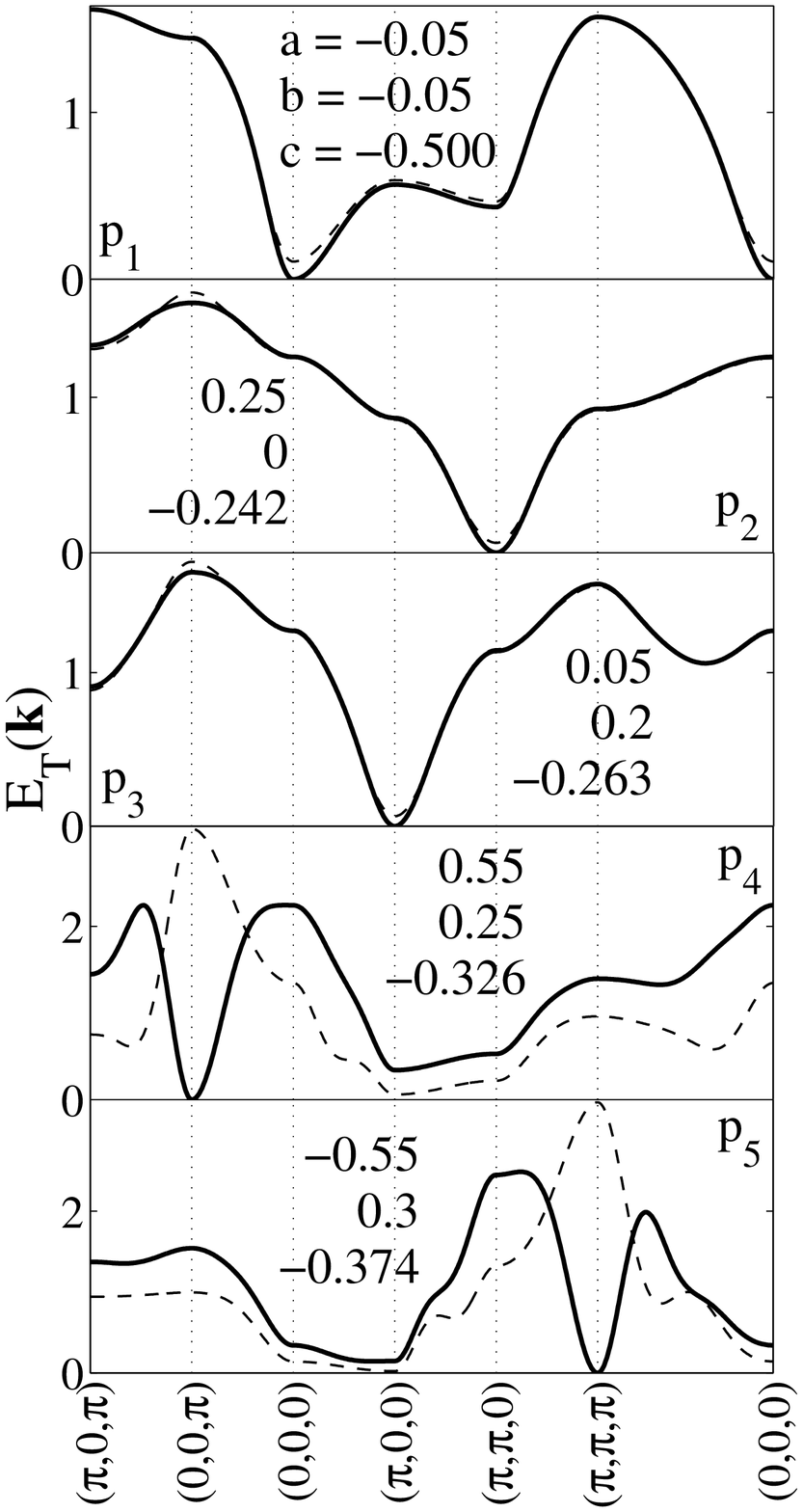,width=0.48\columnwidth}
\psfig{file=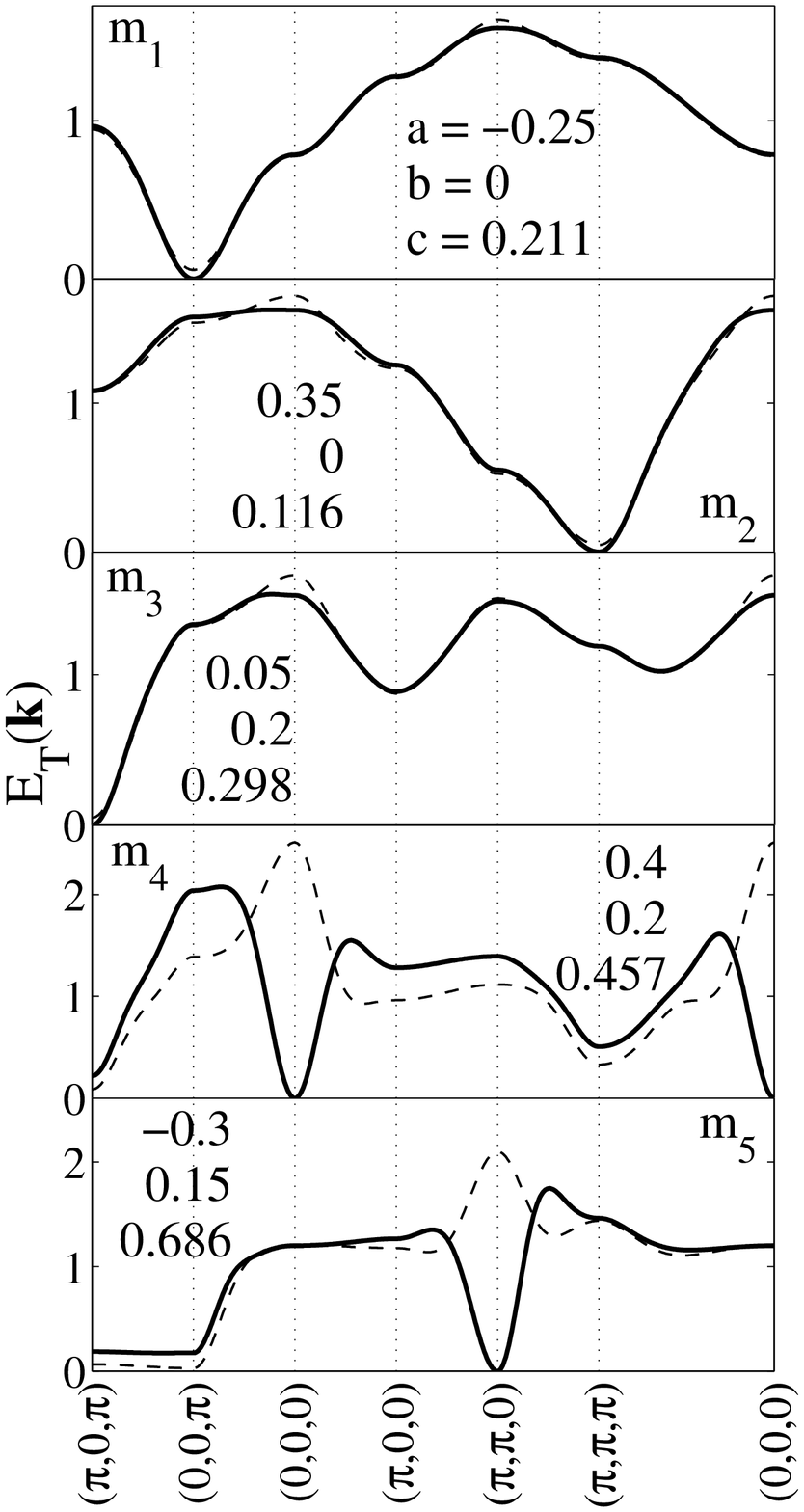,width=0.48\columnwidth}
}
\caption{Left(right) panel: elementary triplet dispersion in the quadrumer
phase for $j_5>(<)0$ at onset of instability. Solid(dashed) line refers
to 4th(3rd) order series expansion. x-axis: path denotes path in
Brillouin zone. Subplot labels $m(p)_{1,...5}$ indicate location of
exchange parameters on instability surface as in fig. \protect{\ref{fig6}} b).
Insets refer to exchange parameters at $m(p)_{1,...5}$.}
\label{fig7}
\vskip-10pt
\end{figure}

Finally we consider the triplet dispersion at critical coupling strengths.
In fig. \ref{fig7} we show $E_{T}({\bf k})$ for wave-vectors ${\bf k}$ along
high-symmetry directions of the Brillouin zone. The exchange parameters have
been selected from the points $p_{1,...5}$ and $m_{1,...5}$ on the instability
surface of fig. \ref{fig6} b). The figure demonstrates a rich variety of
${\bf k}$-dependencies possible. Since Cu$_2$Te$_2$O$_5$Cl$_2$ seems to
order magnetically, inelastic neutron scattering data on the tellurates
would be interesting in order to choose among these dispersion for a set of
exchange constants relevant to the Chlorine system. To check for the
convergence of the
series expansion fig. \ref{fig7} contains both, 3rd- and
4th-order results. On those faces of the instability surface which 
appear as continuous deformations of the 1st order surface of fig.
\ref{fig5}, i.e. for $p(m)_{1,2,3}$ the perturbative result is well converged.
However, within the aforementioned edge-regions, i.e.
for $p(m)_{4,5}$ the convergence is insufficient. In particular the
critical wave-vectors of the instabilities deduced from the 4th-order result
within this region may be an artifact. This remains to be clarified in future
analysis.

To summarize, we have performed a quadrumer series-expansion for a
three-dimensional tetrahedral cluster spin-system using the flow-equation
method. We have have shown our results to incorporate and interpolate
between findings known from previously studied either one- or
two-dimensional quantum spin-systems which are found to be limiting
subsets of our model. We have analyzed the dispersion of the
elementary triplet excitations and the stability of the quadrumer phase
against magnetic ordering. Future studies will have to contrast
this type of ordering against other transitions possible in this cluster
system in order to add more
information towards a complete quantum phase-diagram. We hope that our
results may prompt further investigations of the tellurate compounds
Cu$_2$Te$_2$O$_5$X$_2$, in particular inelastic neutron
scattering studies in order to clarify the relation of the cluster
spin-model to these materials.

It is a pleasure to thank C. Gros, P. Lemmens, F. Mila, and
R. Valenti for comments and helpful discussions.  This research
has been supported by the Deutsche Forschungsgemeinschaft in the
SPP 1073 under Grant BR 1084/1-2.

\appendix
\section{}
To further clarify and connect to other existing analytic approaches, in
this appendix we also list the result to 2nd order in $j_{1,...6}$ for
the triplet dispersion for which we find
\begin{eqnarray}\label{wa1}
\lefteqn{E_{T}({\bf k})=
1+\frac{145 j_2^2}{432}-\frac{31 j_3^2}{864}-\frac{187 j_2 j_4}{432}
-\frac{31 j_4^2}{864}+\frac{8 j_5^2}{27}}&& \nonumber 
\\ \nonumber &&
-\frac{52 j_5 j_6}{27}+\frac{101 j_6^2}{108}+\big(\frac{2 j_3}{3}
+\frac{7 j_3^2}{36}-\frac{1}{9}{{(2 j_2-j_4)}^2}\big)
\\ \nonumber &&
\times \cos (k_x) \cos (k_y)+\big(-\frac{j_2^2}{9}+\big(-\frac{1}{3}+
\frac{5 j_3}{27}\big)(2 j_2-j_4)
\\ \nonumber &&
+\frac{7 j_4^2}{72}\big)(\cos (k_x)+\cos (k_y))+\frac{1}{54} j_3^2 \cos (2 k_x)
\cos (2 k_y)
\\  \nonumber &&
+\big(-\frac{j_2^2}{27}-\frac{j_3^2}{18}+\frac{j_2 j_4}{27}+
\frac{j_4^2}{108}\big) (\cos (2 k_x)+\cos (2 k_y))
\\ \nonumber &&
+\frac{1}{27} j_3 (2 j_2-3 j_4) (\cos (2 k_x) \cos (k_y)+\cos (k_x) \cos (2 k_y))
\\ \nonumber &&
+\big(-\frac{2 j_5^2}{3}-\frac{4 (j_5-j_6)}{3}+\frac{j_6^2}{9}\big)\cos (k_z)
\\ \nonumber &&
+\frac{8}{9} j_3 (j_5-j_6) \cos (k_x) \cos (k_y) \cos (k_z) 
%
%
\\
\nonumber &&
-\frac{4}{9} (2 j_2-j_4) (j_5-j_6) (\cos (k_x)+\cos (k_y)) \cos (k_z)
\\ &&
-\frac{4}{9} {{(j_5-j_6)}^2} \cos (2 k_z)
\;\;.
\end{eqnarray}
In contrast to the 1st-order result, a dependence on combined effective
exchange constants only, i.e. $(2j2-j4)$ and $(j5-j6)$, is absent.
For the case of the 1/5-depleted square lattice, i.e. for $j_1=j_2$ and
$j_3=j_5=j_6=0$, and rewriting (\ref{wa1}) in a form which allows for
direct comparison with the BOT of ref. \cite{Starykh96a} we get 
\begin{eqnarray}\label{wa2}
&& E_{T}({\bf k}) = 1+\frac{59 j_2^2}{144}-\frac{73 j_2 j_4}{144}
-\frac{47 j_4^2}{864}
\\ \nonumber
&& + \big(\frac{1}{3}(j_4 - 2 j_2)-\frac{j_2^2}{9}+\frac{7 j_4^2}{72}\big)
(\cos (k_x)+\cos (k_y))  \\ \nonumber
&& -\frac{1}{18} {{(j_4 -2 j_2)}^2} {{(\cos (k_x)+\cos (k_y))}^2}  \\ \nonumber
&& +\big(\frac{4 j_2^2}{27} -\frac{4 j_2 j_4}{27}+ \frac{2 j_4^2}{27}\big)
({{\cos (k_x)}^2}+{{\cos (k_y)}^2})
\;\;.
\end{eqnarray}
this shows the BOT-dispersion, cited after (\ref{eqn1}), to be correct only
to 1st order. Analogous, for the 'dimerized spin-1 chain sector' of the
tetrahedral chain studied in refs.\cite{Brenig01a,Totsuka02a}, i.e. for
$j_{2,3,4,6}=0$, we may rewrite (\ref{wa1}) into
\begin{eqnarray}\label{wa3}
E_{T}({\bf k}) = &&1+\frac{20 j_5^2}{27}-
\big(\frac{4 j_5}{3}+
\frac{2 j_5^2}{3}\big) 
\cos (k_z)
\nonumber \\
&& -\frac{8}{9}
j_5^2 {{\cos (k_z)}^2}
\;\;.
\end{eqnarray}
Again, the BOT-dispersion is correct to 1st order only. In
ref.\cite{Totsuka02a}, perturbation theory up to 2nd order has been performed
using a very different method than presented here. Therefore it is
satisfying to realize, that (\ref{wa3}) is exactly identical to the
corresponding eqn. (9) in section II.B.2. of ref.\cite{Totsuka02a}.


\begin{thebibliography}{99}
\bibitem{Hase93a}
M. Hase, I. Terasaki, and K. Uchinokura,
Phys.\ Rev.\ Lett. {\bf 70}, 3651 (1993).

\bibitem{Kageyama00a}H. Kageyama, M. Nishi, N. Aso, K. Onizuka,
T. Yoshihama, K. Nukui, K. Kodama, K. Kakurai, and Y. Ueda,
Phys. Rev. Lett. {\bf 84}, 5876 (2000).

\bibitem{Ramirez96a}
P. Schiffer and A.P. Ramirez,
Comments Cond. Mat. Phys. {\bf 18}, 21 (1996).

\bibitem{Ramirez00a}A. P. Ramirez, et al. {\bf all},
Phys. Rev. Lett. {\bf 84}, 2957 (2000).

\bibitem{Mila00a} F. Mila,
Eur. J. Phys. {\bf 21}, 499 (2000),
{\em and references therein}.

\bibitem{Fouet01a}
J.-B. Fouet, M. Mambrini, P. Sindzingre, and C. Lhuillier,
preprint cond-mat/0108070.

\bibitem{Brenig02a}
W. Brenig und A. Honecker,
Phys. Rev. B {\bf 65}, 140407 (2002).

\bibitem{Kawakami00a}
A. Koga and N. Kawakami,
Phys. Rev. B {\bf 63}, 144432 (2001).

\bibitem{Johnsson00a}
M. Johnsson, K.W. T\"ornross, F. Mila, and P. Millet,
Chem. Mater. 12, 2853 (2000).

\bibitem{Lemmens01a}
P. Lemmens, K.-Y. Choi, E.E. Kaul, Ch. Geibel, K. Becker,
W. Brenig, R. Valenti, C. Gros, M. Johnsson, P. Millet, and F. Mila,
Phys. Rev. Lett. {\bf 87}, 227201 (2001).

\bibitem{Gros02a}
C. Gros, {\em et al.}, {\em preprint}
                     
\bibitem{Brenig01a}
W. Brenig and K.W. Becker,
Phys. Rev. B {\bf 64}, 214413 (2001).

\bibitem{Totsuka02a}
K. Totsuka and H.-J. Mikeska,
cond-mat/0207176

\bibitem{Stein97}
J. Stein,
J. Stat. Phys. {\bf 88},  487  (1997).

\bibitem{Mielke98}
A. Mielke,
Eur. Phys. J. B {\bf 5},  605  (1998).

\bibitem{Knetter00}
C. Knetter and G.S. Uhrig,
Eur. Phys. J. B {\bf 13}, 209 (2000).

\bibitem{Wegner94}
F. Wegner,
Ann. Physik {\bf 3},  77  (1994).

\bibitem{HttpRef}
On apparence of this paper additional results are available on request from
the authors. These results are too lengthy to be displayed in written
form but will be made available electronically.

\bibitem{Ueda96}
K. Ueda, H. Kontani, M. Sigrist, and P. A. Lee,
Phys. Rev. Lett. {\bf 76}, 1932 (1996).

\bibitem{Starykh96a}
O. A. Starykh, M. E. Zhitomirsky, D. I. Khomskii, R. R. P. Singh,
and  K. Ueda,
Phys. Rev. Lett. {\bf 77}, 2558 (1996).

\bibitem{Gelfand96a}
M.P. Gelfand, Z. Weihong, R.R.P. Singh, J. Oitmaa, and C.J. Hamer,
Phys. Rev. Lett. {\bf 77}, 2794 (1996). 

\end{thebibliography}
\end{document}